\documentstyle[12pt]{article}
\begin{document}
\begin{center}
$$\;$$
{\Large\bf Dyon--Oscillator Duality}
\footnote{This lecture was presented at the International School
"Symmetries and Integrable Systems" organized by the
Joint Institute for Nuclear Research (Dubna) and the Institute of
Theoretical and Experimental Physics (Moscow) / Dubna, Russia, June 8-11, 1999.}
\end{center}
\begin{center}
{\large V. Ter-Antonyan}
\footnote{e-mail: terant@thsun1.jinr.ru}
\\[3mm]

      {\it Bogoliubov Laboratory of Theoretical Physics,\\
        Joint Institute for Nuclear Research, \\
        Dubna, Moscow Region, 141980, Russia}
\end{center}

\vspace{0.7cm}

\begin{abstract}
The dyon--oscillator duality presented in this lecture
can be treated as a prototype of the Seiberg--Witten duality in nonrelativistic
quantum mechanics. The key statement declares that in some spatial dimensions the
oscillator-like systems are dual to the atoms composed of the electrical charged
particle and dyon, i.e., monopoles
provided by both magnetic and electric charge.
\end{abstract}

\vspace{0.5cm}

\section{Introduction}

The objective of the present lecture is to illustrate the
property of the Schr\"odinger equation which is called here
the dyon--oscillator duality.
The property is in the following. The Sch\"odinger equation for an oscillator
possesses two parameters  -- the energy $E$ and the cyclic frequency $\omega$.
The quantization leads to the constraint $E=\hbar\omega(N+D/2)$ where
$N=0,1,2,\dots$, and $D$ is the dimension of  the configuration
space of the oscillator. If $\omega$ is fixed, then $E$ is quantized
and that is the standard situation. Imagine for a moment that now
$E$ is fixed. Whence, necessarily $\omega$ is quantized, and we are
in a nonstandard situation. The question is whether the nonstandard situation
corresponds to any physics, i.e., whether it is possible  to find such a transformation
that converts the oscillator into a physical system with a coupling constant
$\alpha$, being a function of $E$, and energy $\varepsilon$, depending on $\omega$.
If there  exists such a transformation, we can confirm that
the "nonstandard oscillator" is identical to that physical system.
Below will be shown the validity of the  described picture
for  dimensions $D=1,2,4,8$ and that the final system is a bound system
of charge--dyon (remind, that dyon is the hypothetical particle
introduced by Schwinger which is unlike the Dirac monopole,  endowed with
not just  magnetic but  electric charge as well). As the "standard" and
"nonstandard" regimes are mutually exclusive, the  initial oscillator
and the final "charge--dyon" system are dual to each other, and that
explains the relevancy of the term "dyon--oscillator duality".
Note also that  in the initial system the spectrum is discrete only,
i.e. the particle has just a finite motion
(for such cases it is usually said that we have a model
with confinement). Generally speaking, the spectrum of the final system includes
the discrete spectrum  as well as the continuous one, i.e. in that model there is no
confinement. However, unlike the first model, in the second model
we have monopoles.  There is some analogy between the dyon--oscillator and
the Seiberg--Witten duality, according to which
the gauge theories with strong interactions are equivalent to the theories
having weak interaction on the one hand and topological nontrivial objects,
such as monopoles and dyons are, on the other hand.

\section{Radial Equations}

Let consider the equation
\begin{equation}
\label{f2.1}
\frac{d^2 R}{d u^2}+\frac{D-1}{u}\frac{d R}{d u}-\frac{L(L+D-2)}{u^2}R
+\frac{2\mu}{\hbar^2}
\left(
E-\frac{\mu\omega^2 u^2}{2}
\right)R=0.
\end{equation}
Here $R$ is the radial part of the wave function for the $D$-dimensional oscillator
($D>2$) and $L=0,1,2,\dots$  are the eigenvalues of the  global angular momentum.

Introduce $r=u^2$ and take into account that
$$
\frac{1}{u}\frac{d}{d u}=2\frac{d}{d r}, \quad
\frac{d^2}{d u^2}=2\frac{d}{d r}+4 r\frac{d^2}{d r^2}.
$$
Then,  equation (\ref{f2.1}) transforms into
\begin{equation}
\label{f2.2}
\frac{d^2 R}{d r^2}+\frac{d-1}{r}\frac{d R}{d r}-\frac{l(l+d-2)}{r^2}R
+\frac{2\mu}{\hbar^2}
\left(
\varepsilon+\frac{\alpha}{r}\right)R=0,
\end{equation}
where
\begin{equation}
\label{f2.3}
d=D/2+1, \quad l=L/2,
\end{equation}
\begin{equation}
\label{f2.4}
\varepsilon=-\mu\omega^2/8, \quad
\alpha=E/4.
\end{equation}
This is quite an unexpected result. If $D=4,6,8,10,\dots$, then
$d=3,4,5,6,\dots$, and equation (\ref{f2.2}) is
formally identical to the radial
equation for a $d$-dimensional hydrogen atom (for odd $D>2$ the value of
$d$ is half-integer and so cannot have the meaning of the dimension of the space in
a usual sense). Then, $l$ takes not just integer but half-integer values as well,
and a question arises about the
origin of the  fermion degree of freedom. The answer to the question will
be given later.
Finally, as has been mentioned in the  first section, equations (\ref{f2.1})
and (\ref{f2.2}) are dual to each other and the duality transformation
is $r=u^2$.

Earlier, just the radial part of the wave function of the oscillator was considered.
For the Schr\"odinger equation we must take into account the angular part as well.
Thus, the duality transformation must also include the transformation
of angular variables. If we interpret the change of variables $r=u^2$ as
a mechanism of generation of electric charge, then
(as will be shown later)
the transformation of some angular variables is responsible for the
generation of magnetic charges.

In the next sections, we study dimensions $D=1$ and $D=2$ not considered in
equation (\ref{f2.1}). Then, we analyze the dimensions $D=4$ and
$D=8$.  The  dyon--oscillator duality
is limited to  these four dimensions.
We postpone for a while the discussion of the problem of
selection of the dimensions $D=1,2,4,8$.

\section{1D Coulomb Anyon}

Consider the one-dimensional Schr\"odinger equation
\begin{equation}
\label{f3.1}
\frac{d^2 \Psi}{ du^2}+\frac{2 \mu}{\hbar^2}\left(
E-\frac{\mu\omega^2 u^2}{2}
\right)\Psi=0,
\end{equation}
where $-\infty<u<\infty$. We define a new variable
$$
x=u^2,
$$
and using the identity
$$
\frac{d^2}{d u^2}= 4|x|\left(\frac{d^2}{d x^2}+\frac{1}{2x}\frac{d}{d x}\right)
$$
and setting
\begin{equation}
\label{f3.2}
\Psi=C\, x^{-1/4}\Phi,
\end{equation}
arrive at the equation
\begin{equation}
\label{f3.3}
\frac{d^2 \Phi}{d x^2}+
\frac{2 \mu}{\hbar^2}\left(
\varepsilon_+\frac{\alpha}{|x|}+\frac{\hbar^2}{2\mu}\frac{3}{16 x^2}
\right){\Phi}=0,
\end{equation}
where $\varepsilon$ and $\alpha$ are the same as in (\ref{f2.4}).

Let us introduce the quantity $\nu$ which takes two values: $\nu=1/4$
and $\nu=3/4$, and rewrite the last equation in the form
\begin{equation}
\label{f3.4}
\frac{d^2 \Phi^{(\nu)}}{d x^2}+
\frac{2 \mu}{\hbar^2}\left( \varepsilon-V_c-V_{cs}
\right){\Phi^{(\nu)}}=0,
\end{equation}
where $V_c=-\alpha/|x|$ and $V_{cs}$ is the Calogero--Sutherland potential
$V_{cs}=-\hbar^2\nu(1-\nu)/2\mu x^2$.

In one spatial dimension, a particle moving in the
Calogero--Sutherland potential has a very unusual property.
Unlike the potential $V_{cs}$, the wave function is not invariant
under the replacement $\nu\to (1-\nu)$. It describes a boson for even $\nu$
and a fermion for odd $\nu$. Statistics corresponding to the other values
of $\nu$ is called the fractional statistics, and the system influenced along with
$V_{cs}$ by a potential binding the particle to the center is called
the 1D anyon. So, we have started from the 1D quantum oscillator
and arrived at the 1D Coulomb anyon.

Comparing Eq. (\ref{f3.1}) with Eq. (\ref{f3.4}),
we summarize that there are two alternative possibilities connected with
Eq. (\ref{f3.1}) -- explicit and hidden.
In the first case, the parameter $\omega$ is fixed
($\omega=fix.>0$) and plays a role of the coupling constant, the parameter $E$ is quantized
and has the meaning of energy, and the system is the 1D quantum oscillator. For a
hidden possibility, the parameter $E$ is fixed ($E=fix.>0$), the coupling constant
is equal to $E/4$, $\omega$ is quantized,
the quantity $\varepsilon=-\mu\omega^2/8$ takes
the meaning of energy, and the system is the 1D Coulomb anyon.
Since the 1D Coulomb anyon includes the $1/x^2$ interaction, it pretends to
be a magnetic monopole in one spatial dimension. So, the
anyon--oscillator duality is a prototype of the
dyon--oscillator duality in 1D Quantum Mechanics.

Now we can calculate the energy levels $\varepsilon_n$ and the wave functions
$\Phi_n^{(\nu)}$ in the following way. For energy levels we have
$$
\varepsilon=-\frac{\mu\omega^2}{8}=-\frac{\mu}{8}
\left[\frac{E}{\hbar(2 n +2 \nu)}\right]^2
=-\frac{\mu}{8}
\left[\frac{4\alpha}{\hbar(2 n +2 \nu)}\right]^2=
-\frac{\mu\alpha^2}{2\hbar^2(n+\nu)^2},
$$
where $N=2n+2\nu-1/2$ with $N$ numerating the energy levels $E=\hbar\omega(N+1/2)$
and $n$ being integer and nonnegative.

Consider the wave functions. It follows from (\ref{f3.2})  that
$$
\Phi_n^{(\nu)}=\frac{1}{C}\, x^{1/4}\, \Psi_n^{(\nu)},
$$
where $\Psi_n^{(\nu)}\equiv \Psi$, and therefore
$$
\int\limits_{-\infty}^{\infty}
|\Phi_n^{(\nu)}|^2 \,dx=\frac{1}{|C|^2}\int\limits_{-\infty}^{\infty} x^{1/2}\,
|\Psi_n^{(\nu)}|^2 \, dx.
$$
The integral in the left-hand side is equal to 1, from which it follows that
$$
|C|^2= \int\limits_{-\infty}^{\infty}\, u^2\, |\Psi_N(u)|^2 \,du =
 \, \overline{u^2}=
\frac{2(n+\nu)\hbar}{\mu\omega}.
$$
Thus,
$$
\Phi_n^{(\nu)}=
\frac{(-1)^n}{\sqrt{2}} \sqrt{\frac{\mu\omega}{\hbar (n+\nu)}}\,
x^{1/4} \Psi_n^{(\nu)}
$$
if we choose the phase factor as $(-1)^n$.

Remind that according to the theory of quantum oscillator,
$$
\Psi_N^{(\nu)}=
\left(\frac{\mu\omega}{\pi\hbar}\right)^{1/4} \frac{1}{2^N N!}
e^{-\mu\omega u^2/2} H_N \left(u \sqrt{\frac{\mu \omega}{\hbar}}\right),
$$
where  $H_N(\xi)$ is the Hermite polynomial
$$
H_N(\xi)=(-1)^N e^{\xi^2}\frac{d^N}{d\xi^N} e^{-\xi^2}.
$$
Further, it is known that Hermite polynomials could be expressed in terms
of confluent hypergeometric functions. For our case ($s=0, 1/2$)
$$
H_{2 n+2 s}(z)= (-1)^n \frac{(2n+2s)!}{n!}
(2 z)^{2s} F(-n, 2s+1/2, z^2).
$$
Using the identification $y=x \mu\omega/\hbar$ and the relations $2s+1/2=2\nu$
and $\mu\omega/\hbar=2\mu\alpha/\hbar^2(n+\nu)$, we get
$$
\Phi_n^{(\nu)}=
\sqrt{\frac{\mu\alpha}{\hbar^2}}\frac{1}{2^{n-\nu+1/4}}
\frac{\sqrt{\Gamma(2n+2\nu+1/2)}}{\pi^{1/4}n! (n+\nu)}
\, y^\nu e^{-|y|/2} F(-n, 2\nu, y),
$$
and after taking into account the
duplication formula for Euler's  gamma-function
$$
\Gamma(2z)= 2^{2z-1} \pi^{-1/2} \Gamma(z)\Gamma(z+1/2)
$$
we arrive at the formula
$$
\Phi_n^{(\nu)}=
\frac{\sqrt{\mu\alpha}}{\hbar}
\frac{1}{n+\nu}\,
\frac{1}{\Gamma(2\nu)}\sqrt{\frac{\Gamma(n+2 \nu)}{n!}}\,
y^\nu e^{-|y|/2} F(-n, 2\nu, y).
$$
So, we have two types of 1D Coulomb anyons with $\nu=1/4$ and $\nu=3/4$,
respectively.

\section{Magnetic Vortex}

Now turn to the cyclic oscillator. Here is the first example where
along with the radial variable there appears an angular one.
In the polar coordinates $(u,\varphi)$, where $0\leq u<\infty$,
$0\leq\varphi<2\pi$, the Schr\"odinger  equation takes the form
\begin{equation}
\label{f4.1}
\frac{\partial^2\Psi}{\partial u^2} +\frac1u\frac{\partial\Psi}{\partial u}
+\frac{1}{u^2} \frac{\partial^2\Psi}{\partial \varphi^2}+\frac{2\mu}{\hbar^2}
\left( E-\frac{\mu\omega^2 u^2}{2}\right)\Psi=0.
\end{equation}
Input new variables
\begin{equation}
\label{f4.2}
r=u^2, \quad \phi=2\varphi
\end{equation}
and rewrite equation (\ref{f4.1}) as
\begin{equation}
\label{f4.3}
\frac{\partial^2\Psi}{\partial r^2} +\frac1r\frac{\partial\Psi}{\partial r}
+\frac{1}{r^2} \frac{\partial^2\Psi}{\partial \phi^2}+\frac{2\mu}{\hbar^2}
\left( \varepsilon+\frac{\alpha}{r}\right)\Psi=0.
\end{equation}
where $\varepsilon$ and $\alpha$ are given by expressions (\ref{f2.4}).
Equation (\ref{f4.2}) is identical to the Schr\"odinger equation
for a two-dimensional hydrogen atom; however, $\phi\in [0, 4\pi)$.
Thus, instead of a plane,  we have two-sheeted Riemann surface.
As a consequence,  the single-valuedness condition
$\Psi(r,\phi+4\pi)=\Psi(r,\phi)$ leads to the integer as well as half-integer
eigenvalues for the angular momentum.
The solution of the first type
for $\phi\to(\phi+2\pi)$ does not change the sign, while
the second type solutions under the same transformation change the sign.
Whence, without loss of information we can think of $\Psi(r,\phi)$
defined in the region $0\leq\phi<2\pi$ and having two modifications that
differ from each other by the quantum number $s=0$ or $1/2$. In addition
$\Psi^{(0)}(r,\phi+2\pi)=\Psi^{(0)}(r,\phi)$ and
$\Psi^{(1/2)}(r,\phi+2\pi)=-\Psi^{(1/2)}(r,\phi)$.
We say that these wave functions describe the
system with full inner momentum $s=0$ and $s=1/2$, respectively.

Introduce now the important substitution
\begin{equation}
\label{f4.4}
\Psi^{(s)}(r,\phi)= e^{i s\phi}\,  \overline{\Psi}^{(s)}(r,\phi),
\end{equation}
where $\overline{\Psi}^{(s)}(r,\phi+2\pi)=\overline{\Psi}^{(s)}(r,\phi)$
for $s=0$ as well as for $s=1/2$. From (\ref{f4.3}) and (\ref{f4.4}) it follows
that the function $\overline{\Psi}^{(s)}$ satisfies the equation
\begin{equation}
\label{f4.5}
\frac{\partial^2\overline{\Psi}^{(s)}}{\partial r^2}
+\frac1r\frac{\partial\overline{\Psi}^{(s)}}{\partial r}
+\frac{1}{r^2} \left(\frac{\partial}{\partial\phi}
+ i s\right)^2 \overline{\Psi}^{(s)}
+\frac{2\mu}{\hbar^2}
\left( \varepsilon+\frac{\alpha}{r}\right)\overline{\Psi}^{(s)}=0.
\end{equation}
Now let us clear up to what system there corresponds equation (\ref{f4.5}).
Input the Cartesian coordinates
$$
x_1= r\cos\phi, \quad x_2=r \sin \phi.
$$
As $\partial/\partial\phi=x_1\partial/\partial x_2-x_2\partial/\partial x_1$, then instead of
(\ref{f4.5}) we have
\begin{equation}
\label{f4.6}
\left(\frac{\partial}{\partial x_1}-\frac{i s x_2}{r^2}
\right)^2 \overline{\Psi}^{(s)}+
\left(\frac{\partial}{\partial x_2}+\frac{i s x_1}{r^2}
\right)^2 \overline{\Psi}^{(s)}
+\frac{2\mu}{\hbar^2}
\left( \varepsilon+\frac{\alpha}{r}\right)\overline{\Psi}^{(s)}=0.
\end{equation}
To this equation there corresponds the Hamiltonian
\begin{equation}
\label{f4.7}
\hat H=\frac{1}{2\mu} \left[
\left(\hat{p}_1-\frac{\hbar s x_2}{r^2}\right)^2+
\left(\hat{p}_2+\frac{\hbar s x_1}{r^2}\right)^2
\right]-\frac{\alpha}{r}.
\end{equation}
Input  a vector
$$
\vec A=\frac{g}{r^2}\, (x_2,-x_1)
$$
where $g=\hbar c s/e$ and $e=\sqrt{\alpha}$. As
$$
rot \vec{A}=\frac{\partial A_2}{\partial x_1}
-\frac{\partial A_1}{\partial x_2}= g\left[
\frac{\partial}{\partial x_1}
\left(\frac{\partial}{\partial x_1}\frac1r\right)
+\frac{\partial}{\partial x_2}
\left(\frac{\partial}{\partial x_2}\frac1r\right)
\right]=g\left(\frac{\partial^2}{\partial x_1^2}+
\frac{\partial^2}{\partial x_2^2}\right)\frac1r=
2\pi g \delta(\vec{x}),
$$
then $\vec{A}$ is the vector potential created by the magnetic vortex
of the magnetic charge $g$ and placed in the origin of coordinates.

Now instead of (\ref{f4.7}) we get the Hamiltonian
$$
\hat H=\frac{1}{2\mu} \left(
\hat{p}_\mu-\frac{e}{c} A_\mu\right)^2-
\frac{e^2}{r},
$$
corresponding to the two-dimensional charge--dyon system.
So it is proved that the cyclic oscillator is dual to the charge--dyon system,
being a generalization of a usual two-dimensional hydrogen atom.

Let us discuss the correspondence between the cyclic oscillator and
the charge--dyon system in detail.
It is well-known that in the polar coordinates $(u,\varphi)$ the energy
and wave function of the cyclic oscillator are given by the formulas
$E=\hbar\omega(2 n+|M|+1)$ and
$$
\Psi_{n,M}(u,\varphi)=A_{n,M}\,  u^{|M|}\,  e^{-\mu\omega u^2/\hbar}\,
F\left(-n,|M|+1, \frac{\mu\omega}{\hbar}\,u^2\right)\, e^{i M\varphi},
$$
where $n=0,1,2,\dots$, $M=0, \pm1, \pm2, \dots$. To even and odd wave functions
there correspond even and odd values of $M$. Formally,
the allowance for parity can be  realized by introducing the quantum
numbers $s=0, 1/2$ and $m=0, \pm1, \pm2, \dots$, so that $M=2(m+s)$.

The energy $\varepsilon$ is calculated  similarly
to the one in the previous section
\begin{equation}
\label{f4.8}
\varepsilon=-\frac{\mu e^4}{2\hbar^2 (n+|m+s|+1/2)^2}.
\end{equation}
Next, going over to $r=u^2$ and $\phi=2 \varphi$ we have
$$
\Psi_{n,m}^{(s)}=A_{n,m}^{(s)} r^{|m+s|} e^{-\mu\omega r/\hbar}
F(-n, 2|m+s|+1, \mu\omega r/\hbar) e^{i(m+s)\phi}.
$$
It remains to pass from the two-sheeted Riemann surface $(0\leq\phi< 4\pi)$
to the plane  $(0\leq\phi<2\pi)$, then take into account (\ref{f4.4})
and the last formula with the expression
$\mu\omega/\hbar=2\mu e^2/\hbar^2(n+|m+s|+1/2)$ and,
introducing a new variable $\rho=2\mu e^2r/\hbar^2(n+|m+s|+1/2)$,
write
\begin{equation}
\label{f4.9}
\overline{\Psi}_{n,m}^{(s)} (\rho,\phi)=C_{n,m}^{(s)}\, \rho^{|n+m|}\, e^{-\rho}\,
F(-n, 2|n+m|+1, \rho) \, e^{i m\phi},
\end{equation}
where the normalization constant $C_{n,m}^{(s)}$  is determined by the condition
$$
2\pi\int\limits_{0}^{\infty}\left|
\overline{\Psi}_{n,m}^{(s)}(\rho,\varphi)\right|^2\, r\, dr=1.
$$

Go back to the transformation (\ref{f4.2}) and pass there from the polar coordinates
$(r,\phi)$ to the Cartesian ones $(x_1, x_2)$. Note that $\phi\in[0,4\pi)$.
We have
\begin{eqnarray}
x_1&=& r \cos \phi= u^2 [\cos^2(\phi/2)-\sin^2(\phi/2)]=u_1^2-u_2^2,
\nonumber\\[2mm]
\label{f4.10}
x_2&=& r \sin \phi= 2 u^2 \sin^2(\phi/2)\cos (\phi/2)= 2u_1u_2.
\end{eqnarray}
This transformation is known from celestial mechanics as the Levi--Civita
transformation. In terms of the complex coordinates $z=x_1+i x_2$,
$v=u_1+i u_2$ it takes the form $z=v^2$, i.e., corresponds to the square
of the complex variable $v$. The Levi--Civita transformation
together with the transformations (\ref{f4.2}) and the $\bf{Z_2}$-reduction
compose the duality transformation. Note that
$x=\sqrt{x_1^2+x_2^2}= u^2\equiv u_1^2+u_2^2$. The last expression is known
as the Euler's identity.  Thus, the Levi--Civita transformation is bilinear
coordinate transformation obeying Euler's identity.
This fact is quite noteworthy from the mathematical point of view,
and we will have an opportunity to discuss it.

\section{Charge--Dyon System}

Unlike the two-dimensional space, in four-dimensions
there are several types of "spherical coordinates".
We take the ones used in the theory of symmetrical top
\begin{equation}
\label{f5.1}
u_1+i u_2= u \cos({\beta}/{2}) \, e^{i(\alpha+\gamma)/2}, \quad
u_3+i u_4= u \sin({\beta}/{2}) \, e^{i(\alpha-\gamma)/2}.
\end{equation}
For $u=const$, the position on a sphere is parametrized by
the coordinates $(\alpha,\beta,\gamma)$ that  cover  the sphere
completely  when
$$
\alpha\in [0,2\pi),\quad
\beta\in [0,\pi),\quad
\gamma\in [0,4\pi).
$$
In the coordinates (\ref{f5.1}), the length-element and the Laplacian
are given by
\begin{eqnarray}
dl^2=du^2+\frac{u^2}{4}\left(
d\alpha^2+d\beta^2+d\gamma^2+2\cos\beta \, d\alpha \, d\gamma\right),
\nonumber\\[2mm]
\frac{\partial^2}{\partial u^2_\mu}=\frac{1}{u^3}
\frac{\partial}{\partial u} \left(u^3 \frac{\partial}{\partial u}\right)
-\frac{4}{u^2} \, \hat{J}^2,
\nonumber
\end{eqnarray}
where
\begin{eqnarray}
\hat{J}^2=-\frac{1}{\sin\beta}\frac{\partial}{\partial \beta}
\left(\sin\beta\, \frac{\partial}{\partial \beta}\right)
-\frac{1}{\sin^2\beta}
\left(\frac{\partial^2}{\partial \alpha^2}-2 \cos\beta
\frac{\partial^2}{\partial\alpha\partial\gamma}+\frac{\partial^2}{\partial \gamma^2}
\right).
\nonumber
\end{eqnarray}
Thus, in terms of the coordinates (\ref{f5.1})
the isotropic oscillator is described by the equation
$$
\frac{\partial^2\Psi}{\partial u^2} +\frac{3}{u}\frac{\partial\Psi}{\partial u}
-\frac{4}{u^2}\, \hat{J}^2 \Psi+\frac{2\mu}{\hbar^2}
\left( E-\frac{\mu\omega^2 u^2}{2}\right)\Psi=0.
$$
The operators $\hat{J}^2, \hat{J}_3=-i\partial/\partial\gamma$,
$\hat{J}_{3'}=-i\partial/\partial\alpha$ are mutually
commuting, and their eigenfunction is represented by
the matrix of finite rotations
$$
D_{ms}^j (\alpha, \beta, \gamma)= e^{i m \alpha} d_{m s}^j (\beta) e^{i s\gamma}.
$$
The explicit form of the function $d_{m s}^j (\beta)$ is rather complicated,
it can be found in manuals on Quantum Mechanics.  It is important that
the quantities $j, m$ and $s$  run the values $j=0, 1/2, 1, \dots$ and
$m,s=0, \pm1/2, \pm1,\dots, \pm j$.

Now it is clear that the function $\Psi$ should be of the form
$$
\Psi=R(u) \, D_{m s}^{j} (\alpha,\beta,\gamma).
$$
As the eigenvalues of the operator $\hat{J}^2$ are equal to $j(j+1)$, the radial
function $R(u)$ satisfies the equation
$$
\frac{d^2 R}{d u^2}+\frac{3}{u}\frac{d R}{d u}-\frac{4 j(j+1)}{u^2}R
+\frac{2\mu}{\hbar^2}
\left(
E-\frac{\mu\omega^2 u^2}{2}
\right)R=0.
$$
This equation is solved as follows.
First, use the dimensionless variable $v=a u$ with $a= (\mu\omega/\hbar)^{1/2}$, and rewrite
the last equation as
$$
\frac{d^2 R}{d v^2}+\frac{3}{v}\frac{d R}{d v}-\frac{4 j(j+1)}{v^2}R
+\lambda R -v^2 R=0,
$$
where $\lambda= 2\mu E/\hbar^2 a^2= 2 E/ \hbar \omega$. The next step
is in the passage to the variable $\rho=v^2$ and the equation
$$
\frac{d^2 R}{d \rho^2}+\frac{2}{\rho}\frac{d R}{d v}-\frac{j(j+1)}{\rho^2}R
+\left(\frac{\lambda}{4 \rho}-\frac14\right)R=0.
$$
The analysis of this equation as $\rho\to 0$ and $\rho\to \infty$ verifies
the appropriateness of the substitution
$$
R(u) = \rho^j\, e^{-\rho/2}\, W(\rho)
$$
leading to the equation for a confluent hypergeometric function
$$
\rho \frac{d^2 W}{d\rho^2}+(2 j+2-\rho)\,\frac{d W}{d \rho}-(j+1-\lambda/4)W=0.
$$
A further scenario is usual to any student who masters
the course of Quantum Mechanics. The result is
$$
W=F(j+1-\lambda/4, 2j+2;\rho),
$$
$$
j+1-\lambda/4=-n,\quad n=0,1,2\dots.
$$
Concluding, we receive
\begin{eqnarray}
\label{f5.2}
E_N=\hbar\omega (N+2), \quad N=2n+2j=0,1,2,\dots
\\[2mm]
\label{f5.3}
\Psi=Const (a u)^{2 j}\, e^{-a^2 u^2/2}\, F(-n, 2j +2, a^2 u^2)\,
d_{ms}^j(\beta)\, e^{i m\alpha}\, e^{i s\gamma}.
\end{eqnarray}
For fixed $j$ to the energy level $E_N$ there correspond $(2j+1)^2$
states (degeneracy by $m$ and $s$). As $j=\frac{N}{2}, \frac{N}{2}-1,\dots$,
the total degeneracy for the $N$-th energy level is
$$
g_N=\frac{1}{6}\,(N+1)(N+2)(N+3).
$$
Observe now how
the charge--dyon system
could be obtained from the four-dimensional oscillator.

Using the variable $r=u^2$, we obtain
\begin{eqnarray}
\frac{\partial^2}{\partial u^2_\mu}&=& 4 r \left\{
\frac{1}{r^2}\frac{\partial}{\partial r}
\left(r^2 \frac{\partial}{\partial r}\right)+\frac{1}{r^2}
\left[
\frac{1}{\sin\beta}\frac{\partial}{\partial \beta}\left(
{\sin\beta}\frac{\partial}{\partial \beta}\right)
+\frac{1}{\sin^2\beta}\frac{\partial^2}{\partial\alpha^2}
\right]\right.\nonumber
\\[2mm]
&+&\left.\frac{1}{r^2\sin^2\beta}\left[\frac{\partial^2}{\partial\gamma^2}
-2\cos\beta\frac{\partial^2}{\partial\alpha \, \partial\gamma}\right]
\right\}.\nonumber
\end{eqnarray}
Thus, the Schr\"odinger equation
$$
\frac{d^2 \Psi}{ du_\mu^2}+\frac{2 \mu}{\hbar^2}\left(
E-\frac{\mu\omega^2 u^2}{2}
\right)\Psi=0
$$
gains (in terms of the coordinates (\ref{f5.1})) the form
\begin{eqnarray}
\frac{1}{r^2}\frac{\partial}{\partial r}
\left(r^2 \frac{\partial\Psi}{\partial r}\right)&+&\frac{1}{r^2}
\left[
\frac{1}{\sin\beta}\frac{\partial}{\partial \beta}\left(
{\sin\beta}\frac{\partial\Psi}{\partial \beta}\right)
+\frac{1}{\sin^2\beta}\frac{\partial^2\Psi}{\partial\alpha^2}
\right]\nonumber
\\[2mm]
&+&\frac{1}{r^2\sin^2\beta}\left[\frac{\partial^2\Psi}{\partial\gamma^2}
-2\cos\beta\frac{\partial^2\Psi}{\partial\alpha \, \partial\gamma}\right]
+\frac{2\mu}{\hbar^2}\left(\varepsilon+\frac{e^2}{r}\right)\Psi=0,
\nonumber
\end{eqnarray}
with $e^2=E/4$, $\varepsilon=-\mu\omega^2/8$.

Perform a substitution
\begin{equation}
\label{f5.4}
\Psi(r,\alpha,\beta,\gamma)=\overline{\Psi}^{(s)}(r,\alpha,\beta)\,\,
 e^{is(\alpha+\gamma)},
\end{equation}
where $s$ is any real parameter. It is easy to show that the function
$\overline{\Psi}^{(s)}$ satisfies the equation
\begin{eqnarray}
\frac{1}{r^2}\frac{\partial}{\partial r}
\left(r^2 \frac{\partial\overline{\Psi}^{(s)}}{\partial r}\right)+\frac{1}{r^2}
\left[
\frac{1}{\sin\beta}\frac{\partial}{\partial \beta}\left(
{\sin\beta}\frac{\partial\overline{\Psi}^{(s)}}{\partial \beta}\right)
+\frac{1}{\sin^2\beta}\frac{\partial^2\overline{\Psi}^{(s)}}{\partial\alpha^2}
\right]
\nonumber
\\[2mm]
\label{f5.5}
+\frac{2 i s}{r^2(1+\cos\beta)}\frac{\partial\overline{\Psi}^{(s)}}
{\partial\alpha}
-\frac{2 s^2}{r^2(1+\cos\beta)}\overline{\Psi}^{(s)}
+\frac{2\mu}{\hbar^2}\left(\varepsilon+
\frac{e^2}{r}\right)\overline{\Psi}^{(s)}=0.
\end{eqnarray}
As in the previous sections, the expression for $\varepsilon$  is easily found to be
$$
\varepsilon=-\frac{\mu e^4}{2\hbar^2(n+j+1)^2}.
$$
The wave function $\overline{\Psi}^{(s)}$ is obtained by comparing formulae
(\ref{f5.4}) and (\ref{f5.3}), i.e.,
$$
\overline{\Psi}^{(s)}(r,\alpha,\beta)=C\, \rho^j\,e^{-\rho/2} \, F(-n, 2j+2, \rho)\,
d_{ms}^j(\beta)\, e^{i(m-s)\alpha},
$$
where $\rho=a^2 u^2= 2\mu e^2 r/\hbar^2 (n+j+1)$.

The first line of this equation is nothing but
$\partial^2\overline{\Psi}^{(s)}/\partial x_j^2$ where
$$
x_1+ i x_2= r\sin\beta \, e^{i\alpha}, \quad
x_3= r \cos\beta.
$$
Then,
$$
\frac{2 i s}{r^2(1+\cos\beta)}\frac{\partial}{\partial\alpha}=
\frac{i s}{r^2(1+\cos\beta)} \frac{\partial}{\partial\alpha}+
\frac{\partial}{\partial\alpha}\frac{i s}{r^2(1+\cos\beta)}.
$$
Substituting here the formula
$$
\frac{\partial}{\partial\alpha}=x_1
\frac{\partial}{\partial x_2}-
x_2\frac{\partial}{\partial x_1},
$$
we obtain
\begin{eqnarray}
\frac{2 i s}{r^2(1+\cos\beta)}\frac{\partial}{\partial\alpha}&=&
\frac{i s x_1}{r^2(1+\cos\beta)} \frac{\partial}{\partial x_2}+
\frac{\partial}{\partial x_2}\frac{i s x_1}{r^2(1+\cos\beta)}
\nonumber\\[2mm]
&-&\frac{i s x_2}{r^2(1+\cos\beta)} \frac{\partial}{\partial x_1}-
\frac{\partial}{\partial x_1}\frac{i s x_2}{r^2(1+\cos\beta)}.
\nonumber
\end{eqnarray}
Also note that
$$
\frac{s^2(x_1^2+x_2^2)}{r^4(1+\cos^2\beta)}-
\frac{2 s^2}{r^2(1+\cos\beta)}=-\frac{s}{r^2}.
$$
Then, equation (\ref{f5.5}) can be rewritten as
\begin{eqnarray}
\left(\frac{\partial}{\partial x_1}+
\frac{i s x_2}{r^2 (1+\cos\beta)}\right)^2\overline{\Psi}^{(s)}
&+&\left(\frac{\partial}{\partial x_2}-
\frac{i s x_1}{r^2 (1+\cos\beta)}\right)^2\overline{\Psi}^{(s)}+
\frac{\partial^2 \overline{\Psi}^{(s)}}{\partial x_3^2}
\nonumber
\\[2mm]
&+&\frac{2\mu}{\hbar^2}\left(
\varepsilon+\frac{e^2}{r}-\frac{\hbar^2}{2\mu}\frac{s^2}{r^2}
\right)\overline{\Psi}^{(s)}=0.
\nonumber
\end{eqnarray}
This equation is identical to  the Pauli equation
\begin{eqnarray}
\label{f5.6}
\left(\frac{\partial}{\partial x_j}-i
\frac{e}{c} A_j\right)^2\overline{\Psi}^{(s)}
+\frac{2\mu}{\hbar^2}\left(
\varepsilon+\frac{e^2}{r}-\frac{\hbar^2}{2\mu}\frac{s^2}{r^2}
\right)\overline{\Psi}^{(s)}=0,
\end{eqnarray}
where the vector potential $A_j$ is expressed as follows:
\begin{equation}
\label{f5.7}
\vec{A}=\frac{g \sin\beta}{r(1+\cos\beta)}\, (\sin\alpha, -\cos\alpha, 0)
\end{equation}
with $g=\hbar c s/e$.

The vector potential (\ref{f5.7}) corresponds to the Dirac monopole with
the magnetic charge $g$. So, equation (\ref{f5.6})
describes the motion for a charged particle $e$ in the field
of a dyon with charges $(-e, g)$. The presence of the charge $(-e)$ is indicated by
the term $e^2/r$. The part $(-\hbar^2 s^2/ 2\mu r^2)$ presents
a potential introduced
by Goldhaber with the argument of conservation
of the angular momentum in scattering of a charged particle from
a magnetic monopole. As has been proved by Zwanziger, the addition of such a term
makes a problem, corresponding to equation (\ref{f5.6}), superintegrable.

Thus, we are lucky to  "synthesize" from the isotropic oscillator
the bound charge--dyon system. It remains to clear up one important detail.
As was shown by Dirac, the  introduction of magnetic monopole
in Quantum Mechanics leads to the quantization of an electric charge
$$
e=\frac{\hbar c}{g}\, s, \quad s=0,\pm 1/2, \pm 1, \pm 3/2, \dots.
$$
In our approach, the Dirac quantization condition is deduced from formula (\ref{f5.4}).
The transformation $\gamma\to (\gamma+4 \pi)$ is identical, as it is seen
from the coordinate definition (\ref{f5.1}). Requiring the single-valuedness
for the wave function ${\Psi}(r, \alpha,\beta,\gamma)$, we come to the condition
$s=0, \pm 1/2, \pm 1, \dots$ which, together with  the formula $g=\hbar c s/e$,
leads to the quantization of an electric charge.

So, we have shown that the dyon--oscillator duality is valid for
the four-dimensional oscillator.

Now focus on the duality transformation.
So we have
\begin{eqnarray}
x_1+ i x_2&=& r\sin\beta \, e^{i\alpha}=
2 r\sin(\beta/2)\cos(\beta/2)\,  e^{i \alpha}=
2 r \,  \frac{u_1+i u_2}{u} e^{-i(\alpha+\gamma)/2}
\nonumber\\[2mm]
&\cdot&
\frac{u_3+i u_4}{u} e^{-i(\alpha-\gamma)/2} e^{i\alpha}=
2(u_1+i u_2) (u_3+i u_4),
\nonumber\\[2mm]
x_3&=& r\cos\beta= r [\cos^2(\beta/2)-\sin^2(\beta/2)]=r\,
\frac{u_1^2+u_2^2}{u^2}-
r\, \frac{u_3^2+u_4^2}{u^2}
\nonumber\\[2mm]
&=&u_1^2+u_2^2-u_3^2-u_4^2,
\nonumber
\end{eqnarray}
or, otherwise,
\begin{eqnarray}
x_1&=&2 (u_1 u_3-u_2 u_4),
\nonumber\\[2mm]
x_2&=&2 (u_1 u_4+u_2 u_3),
\nonumber\\[2mm]
x_3&=& u_1^2+u_2^2-u_3^2-u_4^2.
\nonumber
\end{eqnarray}
This bilinear transformation satisfies Euler's condition
$r=u^2$ and is called the Kustaanheimo--Stiefel transformation.
It corresponds to the mapping $\rm I\!R^4(\vec{u})\to \rm I\!R^3(\vec{x})$
that, along with the formula
$$
\gamma=\frac{i}{2} \,\, ln\left\{\frac{u_1+i u_2}{u_1-i u_2}\,
\frac{u_3-i u_4}{u_3+i u_4}\right\}
$$
and the ansatz $\Psi\to\overline{\Psi}^{(s)}$,  composes
the duality transformation.

\section{Magic Numbers}

Let us answer the question
why the dyon--oscillator duality is valid just for
the oscillators with the configuration spaces of dimensions $D=1,2 ,4, 8$.
We have already mentioned
that the duality transformation  must satisfy the Euler's identity
\begin{equation}
\label{f6.1}
(u_1^2+u_2^2+\dots +u_D^2)^2=x_1^2+x_2^2 +\dots +x_d^2,
\end{equation}
where $d=1$ for $D=1$ and $d=D/2+1$ for $D>1$.
It was proved by Hurwitz that in the cases of  $x_i$ being
a bilinear combination of $u_i$, the identity
\begin{equation}
\label{f6.2}
(u_1^2+u_2^2+\dots +u_D^2)^2=x_1^2+x_2^2 +\dots +x_D^2
\end{equation}
is true for  $D=1,2,4,8$. These magic numbers are directly related to
the existence of the four fundamental algebraic structures:
real numbers, complex numbers, quaternions and  octonions.
Putting  in (\ref{f6.2}) $x_{d+1}=x_{d+2}=\dots=x_D=0$, we come to  (\ref{f6.1}).

\section{Hurwitz Transformation}

The question arises, of how to find a transformation converting
$\rm I \! R^8(u)$ into $\rm I \! R^5(x)$, i.e. the transformation with
the last of the magic numbers presented above.
Begin to write down the transformation in the form
$$
x= H(u; D)\, u.
$$
Here $D$ is the dimension of the space, $H$ is the matrix $D\times D$
with the elements $u_\mu$, and $x, u$  are the $D$-dimensional columns
composed from $x_j, u_\mu$ and, possibly, zeroes.
So for the Levi--Civita and Kustaanheimo--Stiefel transformations, we have
$$
\left|
\begin{array}{c}
x_1\\x_2
\end{array}
\right|=\left|
\begin{array}{cc}
u_1&-u_2\\u_2&u_1
\end{array}
\right|
\left|
\begin{array}{c}
u_1\\u_2
\end{array}
\right|,
$$
$$
\left|
\begin{array}{c}
x_1\\x_2\\x_3\\0
\end{array}
\right|=\left|
\begin{array}{cccc}
u_3&-u_4&u_1&-u_2\\
u_4&u_3&u_2&u_1\\
u_1&u_2&-u_3&-u_4\\
u_2&-u_1&-u_4&u_3
\end{array}
\right|
\left|
\begin{array}{c}
u_1\\u_2\\u_3\\u_4
\end{array}
\right|.
$$
The matrices $H(u;2)$ and $H(u;4)$ have the property
$$
H(u;2)\, H^T(u;2)=u^2 E(2),\quad
H(u;4)\, H^T(u;2)=u^2 E(4),
$$
where $"T"$ means the sign of transposition, $E(2)$ and $E(4)$
are the unit matrices. Due to these properties the Euler's identities are
fulfilled.  Now, one can easily deduce
that the transformation $\rm I\!R^8(\vec{u})\to\rm I\!R^5(\vec{x})$
must take the form
$$
\left|
\begin{array}{c}
x_0\\x_1\\x_2\\x_3\\x_4\\0\\0\\0
\end{array}
\right|=\left|
\begin{array}{cccccccc}
u_0&u_1&u_2&u_3&-u_4&-u_5&-u_6&-u_7\\
u_4&u_5&-u_6&-u_7&u_0&u_1&-u_2&-u_3\\
u_5&-u_4&u_7&-u_6&-u_1&u_0&-u_3&u_2\\
u_6&u_7&u_4&u_5&u_2&u_3&u_0&u_1\\
u_7&-u_6&-u_5&u_4&u_3&-u_2&-u_1&u_0\\
u_1&-u_0&u_3&-u_2&u_5&-u_4&u_7&-u_6\\
u_2&-u_3&-u_0&u_1&-u_6&u_7&u_4&-u_5\\
u_3&u_2&-u_1&-u_0&-u_7&-u_6&u_5&u_4\\
\end{array}
\right|
\left|
\begin{array}{c}
u_0\\u_1\\u_2\\u_3\\u_4\\u_5\\u_6\\u_7
\end{array}
\right|.
$$

Whence it follows that
\begin{eqnarray}
x_0&=&u_0^2+u_1^2+u_2^2+u_3^2-u_4^2-u_5^2-u_6^2-u_7^2,
\nonumber\\[2mm]
x_1&=&2 \,(u_0 u_4+u_1 u_5-u_2 u_6-u_3 u_7),
\nonumber\\[2mm]
\label{f7.1}
x_2&=&2 \,(u_0 u_5-u_1 u_4+u_2 u_7-u_3 u_6),
\\[2mm]
x_3&=&2 \,(u_0 u_6+u_1 u_7+u_2 u_4+u_3 u_5),
\nonumber\\[2mm]
x_4&=&2 \,(u_0 u_7-u_1 u_6-u_2 u_5+u_3 u_4).
\nonumber
\end{eqnarray}
It is easy to prove that for the matrix $H(u;8)$ there is a condition
$$
H(u; 8)H^T(u;8)=u^2 E(8)
$$
that guarantees the validity of Euler's identity.

Adding to (\ref{f7.1}) the transformations
\begin{eqnarray}
\alpha_T&=&\frac{i}{2}\, \ln \,\frac{(u_0+i u_1)(u_2-i u_3)}{(u_0-i u_1)(u_2+i u_3)},
\nonumber\\[2mm]
\label{f7.2}
\beta_T&=&2 \arctan \left(\frac{u_0^2+u_1^2}
{u_2^2+u_3^2}\right)^{1/2},
\\[2mm]
\gamma_T&=&\frac{i}{2}\, \ln \,\frac{(u_0-i u_1)(u_2-i u_3)}
{(u_0+i u_1)(u_2+i u_3)},
\nonumber
\end{eqnarray}
we obtain a transformation converting $\rm I\!R^8$ to the direct product
$\rm I\!R^5 \otimes{\bf S^3}$ of the space  $\rm I\!R^5(\vec{x})$
and a three-dimensional sphere ${\bf S^3}(\alpha_T,\beta_T,\gamma_T)$.

\section{Yang--Coulomb Monopole}

In the coordinates (\ref{f7.1})-(\ref{f7.2})the eight-dimensional isotropic oscillator
is described by the equation
\begin{equation}
\label{f8.1}
\frac{1}{2\mu}\left(
-i\hbar\frac{\partial}{\partial x_j}-\hbar A_j^2\hat{T}_a
\right)^2\Psi+
\frac{\hbar^2}{2\mu r^2}\hat{T}^2\Psi
-\frac{e^2}{r}\Psi=\varepsilon \Psi
\end{equation}
where $\varepsilon$ and $e^2$ are defined as usual.
The operators $\hat{T}_a$ are the generators of the $SU(2)$ group.
In the coordinates $(\alpha_T,\beta_T,\gamma_T)$ they are parametrized as
follows:
\begin{eqnarray}
\hat{T}^1&=&
i\left(\cos\alpha_T\, \cos\beta_T\frac{\partial}{\partial \alpha_T}
+\sin\alpha_T\frac{\partial}{\partial \beta_T}
-\frac{\cos\alpha_T}{\sin\beta_T}\, \frac{\partial}{\partial \gamma_T}
\right),
\nonumber\\[2mm]
\hat{T}^2&=&
i\left(\sin\alpha_T\, \cot\beta_T\frac{\partial}{\partial \alpha_T}
-\cos\alpha_T\frac{\partial}{\partial \beta_T}
-\frac{\sin\alpha_T}{\sin\beta_T}\, \frac{\partial}{\partial \gamma_T}
\right),
\nonumber\\[2mm]
\hat{T}^3&=&
-i\frac{\partial}{\partial \alpha_T}.
\nonumber
\end{eqnarray}
The five-dimensional vectors $\vec{A}^a$ are given by the expressions
\begin{eqnarray}
\vec{A}^1&=&\frac{1}{r(r+x_0)}\,(0,-x_4, -x_3, x_2, x_1),
\nonumber\\[2mm]
\vec{A}^2&=&\frac{1}{r(r+x_0)}\,(0,x_3, -x_4, -x_1, x_2),
\nonumber\\[2mm]
\vec{A}^3&=&\frac{1}{r(r+x_0)}\,(0,x_2, -x_1, x_4,-x_3).
\nonumber
\end{eqnarray}
Each term of the triplet $A_j^a$ coincides with the vector potential of a $5D$
Dirac monopole with a unit topological charge and with the line of singularity
along the  nonpositive $x_0$ semiaxis.
The vectors $A_j^a$  are orthogonal to  each other
$$
A_j^a A_j^b=\frac{1}{r^2}\, \frac{r-x_0}{r+x_0}\,\delta_{ab}
$$
and to the vector  $\vec{x}=(x_0, x_1, x_2, x_3, x_4)$ as well.

We see that equation (\ref{f2.4}) describes the charge--dyon system with
$SU(2)$ monopoles which we call the Yang--Coulomb monopole (YCM). The YCM is
defined as a five-dimensional system composed of the Yang monopole ($A_j^a$)
of the topological charge $+1$ and the
particle  of the isospin $(\hat{T}_a)$. Both the monopole and particle are
also assumed to have electric charges of the opposite signs. Thus, the
monopole--particle coupling is realized not only by the $SU(2)$ gauge field but also
by the Coulomb interaction. At large distances the Coulomb structure becomes
immaterial and YCM seems to be a pure Yang monopole.
The YCM is a unique example
of an integrable non-Abelian system. The $SO(6)$ group is a group of hidden symmetry
of YCM which can be used for calculation of the energy spectrum of YCM by
an algebraic method.

After quite complicated calculations, which are  omitted here, we can reduce
equation (\ref{f8.1}) to the form
\begin{equation}
\label{f8.2}
\left(\Delta_5-\frac{4}{r(r+x_0)}\hat{L}\hat{T}
-\frac{2}{r(r+x_0)}\hat{T}^2\right)\Psi
+\frac{2\mu}{\hbar^2}\left(\varepsilon+\frac{e^2}{r}\right)\Psi=0,
\end{equation}
where
\begin{eqnarray}
\hat{L}_1&=&\frac{i}{2}[D_{41}(x)+D_{32}(x)], \nonumber
\\[2mm]
\hat{L}_2&=&\frac{i}{2}[D_{13}(x)+D_{12}(x)], \nonumber
\\[2mm]
\hat{L}_3&=&\frac{i}{2}[D_{12}(x)+D_{34}(x)] \nonumber
\end{eqnarray}
with
$$
D_{ij}=-x_i\frac{\partial}{\partial x_j}
+x_j\frac{\partial}{\partial x_i}.
$$
We see that equation (\ref{f8.2}) contains the LT-coupling term demonstrating that
we have no way to separate the wave function dependence on $\rm I\!R^5$
and $\bf{S}^3$.

In  $\rm I\!R^5$ we introduce the hyperspherical coordinates $r\in [o,\infty)$,
$\theta\in[0,2\pi]$, $\alpha\in[0, 2\pi)$, $\beta\in [0,\pi]$ and $\gamma\in
[0, 4\pi)$ according to the relations
\begin{eqnarray}
x_0&=& r\cos\theta,
\nonumber
\\[2mm]
x_1+i x_2&=& r\sin\theta\cos\frac{\beta}{2} e^{i\frac{\alpha+\gamma}{2}},
\nonumber
\\[2mm]
x_3+i x_4&=& r\sin\theta\sin\frac{\beta}{2} e^{i\frac{\alpha-\gamma}{2}},
\nonumber
\end{eqnarray}
and rewrite equation (\ref{f8.2}) as
\begin{equation}
\label{f8.3}
\left(\Delta_{r\theta}-\frac{\hat{L}^2}{r^2\sin^2(\theta/2)}
-\frac{\hat{J}^2}{r^2\cos^2(\theta/2)}\right)\Psi
+\frac{2\mu}{\hbar^2}\left(\varepsilon+\frac{e^2}{r}\right)\Psi=0,
\end{equation}
where $\hat{J}_a=\hat{L}_a+\hat{T}_a$ and
$$
\Delta_{r\theta}=
\frac{1}{r^4}\,
\frac{\partial}{\partial r}\left(
r^4\, \frac{\partial}{\partial r}\right)+\frac{1}{r^2\sin^3\theta}
\frac{\partial}{\partial \theta}\left(\sin^3\theta
\frac{\partial}{\partial \theta}\right).
$$
We introduce the separation ansatz
$$
\Psi=\Phi(r,\theta) G(\alpha,\beta,\gamma; \alpha_T,\beta_T,\gamma_T),
$$
where $G$ are the eigenfunctions of $\hat{L}^2, \hat{T}^2$ and $\hat{J}^2$
with the eigenvalues $L(L+1)$, $T(T+1)$ and $J(J+1)$.

Because of the  LT--interaction, we seek the function $G$ in the form
$$
G=\sum\limits_{M=m+t}\, (JM|L,m;T,t)\, D_{m m'}^L (\alpha,\beta,\gamma)\,
D_{t t'}^T (\alpha_T,\beta_T,\gamma_T),
$$
where  $(JM|L,m;T,t)$ are the Clebsch--Gordan coefficients.

Let us take the function $\Phi(r,\theta)$ in the form
$$
\Phi(r,\theta)= R(r) Z(\theta).
$$
Equation (\ref{f8.3}) is then separated into
\begin{equation}
\label{f8.4}
\frac{1}{\sin^3\theta}
\frac{d}{d \theta}\left(\sin^3\theta
\frac{d Z}{d \theta}\right)
-\frac{2L(L+1)}{1-\cos\theta} \, Z
-\frac{2J(J+1)}{1+\cos\theta} \, Z+\lambda(\lambda+3) Z=0
\end{equation}
and a purely radial equation
\begin{equation}
\label{f8.5}
\frac{1}{r^4}\,
\frac{d }{d r}\left(
r^4\, \frac{d R}{d r}\right)-\frac{\lambda(\lambda+3)}{r^2}
+\frac{2\mu}{\hbar^2}\left(\varepsilon+\frac{e^2}{r}\right) R=0
\end{equation}
with the separation constant $\lambda(\lambda+3)$ being equal to the
nonnegative
eigenvalues of the global angular momentum.

In equation (\ref{f8.4}) it is convenient to change the variable as $y=(1-\cos\theta)/2$
and set
$$
Z(y)=y^L\,(1-y)^J\, W(y).
$$
Substituting this into equation (\ref{f8.4}) we obtain the hypergeometric equation
$$
y(1-y)\frac{d^2 W}{d y^2} +[c-(a+b+1) y]\frac{d W}{d y} -ab W=0,
$$
where $a=-\lambda+L+J$, $b=\lambda+L+ J+3$, $c=2L+2$.

Thus, we find that
$$
Z(\theta)=(1-\cos\theta)^L (1+cos\theta)^L
F\left(-\lambda+J+L,\lambda+J+L+3, 2L+2;\frac{1-\cos\theta}{2}\right).
$$
The solution behaves well at $\theta=\pi$ if the series $F$ terminates, that is
$$
-\lambda+J+L=-n_\theta,
$$
with $n_\theta=0,1,2,\dots$.

Let us now consider the radial equation and introduce the function
$$
f(r)=e^{-k r} r^{-\lambda} R(r).
$$
It can easily be verified that the equation for $f(r)$ has the form of the confluent
hypergeometric equation
$$
z\frac{d^2 f}{d z^2} +(c-z)\frac{d f}{d z} -a f=0,
$$
where $z=2 k r, k=\sqrt{-2\mu\varepsilon/\hbar^2}$, $c=2\lambda+4$,
$a=\lambda+2-1/k r_0$, $r_0=\hbar^2/me^2$. For the bound state solutions
($\varepsilon<0$), we have
$$
\lambda+2-1/k r_0=-n_r,\quad n_r=0,1,2,\dots;
$$
therefore,
$$
\varepsilon_N^T=-\frac{m e^4}{2\hbar^2(N/2+2)^2},
$$
where $N=2(n_r+\lambda)=2(n_r+n_\theta+J+L)$.

For fixed $T$, the energy levels $\varepsilon_N^T$ do not depend on $L,J$ and
$\lambda$, i.e., they are degenerate. The total degeneracy is
$$
g_N^T=(2T+1)\, \sum_{\lambda}\sum_{L} \ (2L+1)\sum_{J}(2J+1).
$$
After some tedious calculation we finally obtain
$$
g_N^T=\frac{1}{12} (2T+1)^2
\left(\frac{N}{2}-T+1\right)
\left(\frac{N}{2}-T+2\right)
\left\{
\left(\frac{N}{2}-T+2\right)
\left(\frac{N}{2}-T+3\right)+2T(N+5)
\right\}.
$$
For $T=0$ and $N=2n$ (even) the right-hand side of the last formula
is equal to $(n+1) (n+2)^2(n+3)/12$ -- that is, to the degeneracy of pure Coulomb
levels. Further, we have $T=0,1,\dots, N/2$ for even $N$ and
$T=1/2,3/2,\dots, N/2$ for odd $N$. Therefore,
$$
g_N=\sum^{N/2}_{T=0, \frac12} g_N^T=\frac{(N+7)!}{7! N!}
$$
i.e., we obtain the degeneracy of the energy levels for the 8D isotropic
quantum oscillator.

Formulae (\ref{f7.1}) and (\ref{f7.2}) represent the duality transformation
mapping the 8D quantum oscillator into charge--dyon
system with the $SU(2)$ monopole.

\section{Oscillator-like Systems}

We have considered above the dyon--oscillator duality. This type of duality
is valid not only for the $1D, 2D, 4D$ and $8D$ oscillators,
but also for oscillator-like systems with the potentials
$$
V(u^2)=C_0+C_2 u^2 +W(u^2),
$$
where $W(u^2)$ has the form
$$
W(u^2)=\sum\limits_{n=2}^{\infty} C_{2n} u^{2n}
$$
For such modified potentials the ansatz (\ref{f2.4}) can be rewritten as
$$
\varepsilon=-\frac{C_2}{4}, \quad e^2=\frac{E-C_0}{4}
$$
Thus, the value of the function $V(u^2)$ at $u^2=0$ contributes to the
Coulomb coupling constant $e^2$. It is also easy to verify that the left-hand side
of equation (\ref{f8.5}) develops the additional term $(-W(r)/4 r)$.

\vspace{1cm}

\section{Exercises}

\vspace*{0.01cm}

$\bullet$
 Find the energy levels and normalized wave functions
for states of the particle placed in the field $V(x)=-\alpha/|x|-\hbar^2\nu(\nu-1)/
2\mu x^2$, where $x\in(-\infty,\infty)$ and $\nu\not = 0,1/2$ (see Ref. [50]).

\vspace{0.2cm}

\noindent
$\bullet$ Prove that the 3D oscillator with coordinates confined by the 2D half-up cone
for an angle of $\pi/6$ is dual to the 2D charge--dyon system obeying
fractional statistics. Find the duality transformation (see Ref. [48]).

\vspace{0.2cm}

\noindent
$\bullet$ Calculate the length-element $d l^2$, metric tensor $g_{\mu\nu}$ and
the Laplace operator $\partial^2/\partial u^2_\mu$ in the coordinates
(\ref{f5.1}).

\vspace{0.2cm}

\noindent
$\bullet$ Compute the integrals of motion for the
3-dimensional charge--dyon system,
transforming for $g=0$ into the operator of orbital momentum and
Runge--Lenz operator (see Ref. [45]).

\vspace{0.2cm}

\noindent
$\bullet$ Prove that the Goldhaber correction in the Hamiltonian
of the 3-dimensional charge--dyon system is identical to the interaction
$\vec{\mu}\vec{B}$ of  the magnetic momentum $\vec{\mu}$ of a particle  with
the magnetic field $\vec{B}$ (see Ref. [45]).

\vspace{0.2cm}

\noindent
$\bullet$ Solve the Schr\"odinger equation for
the 3-dimensional charge--dyon system in the parabolic
coordinates $x_1=\sqrt{\xi\eta}\cos\varphi$
$x_2=\sqrt{\xi\eta}\sin\varphi$, $x_3=(\xi-\eta)/2$ (see Ref. [49]).

\vspace{0.2cm}

\noindent
$\bullet$ Compute the expansion coefficients of the parabolic basis of
the 3-dimensional charge--dyon system  in terms of its spherical basis (see
Ref.  [49]).

\vspace{0.2cm}

\noindent
$\bullet$ Prove that the transformation (\ref{f7.1}) converts the Schr\"odinger
equation for the $8D$ oscillator  into equation (\ref{f8.1}).

\vspace{0.2cm}

\noindent
$\bullet$ Show that equation (\ref{f8.1}) can be transformed
into (\ref{f8.2}).

\vspace{0.2cm}

\noindent
$\bullet$ Calculate the length-element $d l^2$, metric tensor $g_{ij}$ and Laplace operator
$\partial^2/\partial x_j^2$
in the coordinates $(r,\theta,\alpha,\beta,\gamma)$.

\vspace{1cm}

{\large\bf Acknowledgment}

\vspace{0.5cm}

I would like to thank  the organizers of the school
for the occasion to give a talk  for so
young and active audience.

Also I am grateful to Yeranuhi Hakobyan for her help in preparing
this lecture.

\vspace{2cm}

\vspace{1cm}

{\Large \bf References}

\vspace{1cm}

{\it Duality in QFT}:

\noindent
[1]  N. Seiberg and E. Witten, Nucl. Phys. {\bf B 431}, 484 (1994).

\noindent
[2] Jeffrey A. Harvey, "Magnetic Monopoles, Duality, and Supersymmetry",
EFI-96-06,hep-th/9603-86.

\vspace{0.5cm}

{\it Dirac Monopoles and Dyons}:

\noindent
[3] P.A.M. Dirac, Proc. Roy. Soc. {\bf A 133}, 69 (1931).

\noindent
[4] J. Schwinger, Science, {\bf 165}, 757 (1969).

\noindent
[5] T.T.Wu and C.N. Yang, Phys.Rev. {\bf D 12}, 3845 (1975).

\vspace{0.5cm}

{\it Charge--Dyon and Dyon--Dyon Systems}:

\noindent
[6] A.S. Goldhaber, Phys. Rev. {\bf B 140}, 1407 (1965).

\noindent
[7] D. Zwanziger, Phys. Rev. {\bf 176}, 1480 (1968).

\vspace{0.5cm}

{\it One-Dimensional Anyons}:

\noindent
[8] A.P. Polychronakos, Nucl. Phys. {\bf B 324}, 597 (1989).

\noindent
[9] A.P. Balachandran, Int. J. Mod. Phys. {\bf B 5}, 2585 (1991).

\noindent
[10] C.A. Anerisis and A.P. Balachandran, Int. J.Mod.Phys. {\bf A 6}, 4721 (1991).

\noindent
[11] S. Isakov, Int. J. Mod. Phys.  {\bf A 9}, 2563 (1994).

\vspace{0.5cm}

{\it Two-Dimensional Anyons}:

\noindent
[12] J.M. Leinaas and Myrheim, Nuovo Ciemento {\bf B 37}, 1 (1977).

\noindent
[13] F. Wilczek, Phys. Rev. Lett. {\bf 44}, 957 (1982).

\noindent
[14] R. Jackiw, Ann. Phys. {\bf 201}, 83 (1990).

\noindent
[15] G.S. Canright and S.M. Girvin, Science {\bf 247}, 1197 (1990).

\noindent
[16] S. Forte, Rev. Mod. Phys. {\bf 64}, 193 (1992).

\noindent
[17] Alberto Lenda, "Anyons: Quantum Mechanics of Particles with Fractional
Statistics" (Springer-Vrelag, Berlin Heidelberg, 1992).

\vspace{0.5cm}

{\it Group Theoretical Approach to 3D Anyons}:

\noindent
[18] M.S. Plyushchay, Phys. Lett. {\bf B 262}, 71 (1991).

\noindent
[19] M.S. Plyushchay, Phys. Lett. {\bf B 362}, 54 (1991).

\noindent
[20] M.S. Plyushchay, Nucl. Phys. {\bf B 491}, 619 (1997).

\noindent
[21] M.S. Plyushchay, Mod. Phys. Lett. {\bf A 12}, 1153 (1997).

\vspace{0.5cm}

{\it Yang Monopole}:

\noindent
[22] C.N. Yang, J. Math. Phys. {\bf 19}, 320; 2622 (1978).

\noindent
[23] M. Minami, Proc. Theor. Phys. {\bf 63}, 303 (1980).

\vspace{0.5cm}

{\it Hurwitz Transformation}:

\noindent
[24] A. P. Chen and M. Kibler, Phys. Rev.  {\bf A 31}, 3960 (1985).

\noindent
[25] D. Lambert and M. Kibler, J. Phys. {\bf A 21}, 307 (1988).

\noindent
[26] M. Kibler, P. Winternitz, J. Phys. {\bf A 21}, 1787 (1988).

\noindent
[27] L. Davtyan, A. Sissakian, V. Ter-Antonyan, J. Math. Phys. {\bf 36}, 1 (1995).

\noindent
[28] A. Sissakian, V. Ter-Antonyan, "The Structure of the Hurwitz Transformation".
In Proceedings of the International Workshop "Finite Dimensional Integrable Systems",
191, Dubna, 1995.

\noindent
[29] L. Mardoyan, A. Sissakian, V. Ter-Antonyan, "The Eulerian Parametrization
of the Hurwitz Transformation". In Proceedings of the International Workshop
"Finite Dimensional Integrable Systems", 137, Dubna, 1995.

\vspace{0.5cm}

{\it Oscillator and Kepler Problem}:

\noindent
[30] T. Iwayi, Y. Uwano, J. Math. Phys. {\bf 27}, 1523 (1986).

\noindent
[31] M. Kibler, A. Ronveaux, T. Negadi, J. Math. Phys. {\bf 27}, 1541 (1986).

\noindent
[32] T. Iwayi, Y. Uwano, J. Phys. {\bf A 21}, 4083 (1988).

\noindent
[33] A. Inomata, G. Junker, R. Wilson, Found. Phys. {\bf 23}, 1075 (1993).

\noindent
[34] T. Iwai, T. Sunako, J. Geom. Phys. {\bf 20}, 250 (1996).

\noindent
[35] Gao-Jian Zeng, Steng-Mei Ao, Xiang- Sheng Wu and Ka-Lin-Su, Int. J. Theor. Phys. {\bf 37},
2463 (1998).

\noindent
[36] M.V. Pletyukhov and E. A. Tolkachev, Rep. Math. Phys.
{\bf 43}, 303  (1999).

\noindent
[37] M.V. Pletyukhov and E. A. Tolkachev, J. Phys. {\bf A 32},
L249  (1999).

\noindent
[38] Toshiniro Iwai, J. Geom. Phys. {\bf 7}, 507 (1990).

\noindent
[39] I. Mladenov and V. Tsanov, J.Phys. {\bf A 20}, 5865 (1987).

\noindent
[40] L. Davtyan, L. Mardoyan, G. Pogosyan, A. Sissakian, V. Ter-Antonyan,
J. Phys.  {\bf A 20}, 6121 (1987).

\noindent
[41] M.V. Pletyukhov and E.A. Tolkachev, J. Math. Phys. {\bf 40},
93 (1999).

\vspace{0.5cm}

{\it Singular D-Dimenional Oscillator}:

\noindent
[42]  B. Sutherland, J. Math. Phys. {\bf 12}, 246 (1971).

\noindent
[43]  F. Calogero, J. Math. Phys. {\bf 12}, 419 (1971).

\noindent
[44] Ye. Hakobyan, G. Pogosyan and A. Sissakian, Phys. Atom. Nucl. {\bf 61},
1762 (1998).

\vspace{0.5cm}

{\it Dyon--Oscillator Duality (Abelian Monopole)}:

\noindent
[45] A. Nersessian, V. Ter-Antonyan, Mod. Phys. Lett. {\bf A 9}, 2431 (1994).

\noindent
[46] A. Nersessian, V. Ter-Antonyan, Mod. Phys. Lett. {\bf A 10}, 2633 (1995).

\noindent
[47] A. Nersessian, V. Ter-Antonyan, M. Tsulaia, Mod. Phys. Lett. {\bf A 11},
1605 (1996).

\noindent
[48] A. Maghakian, A. Sissakian, V. Ter-Antonyan, Phys. Lett. {\bf A 236}. 5 (1997).

\noindent
[49] L. Mardoyan, A. Sissakian, V. Ter-Antonyan, Int. J. Mod. Phys. {\bf A 12},
237 (1997).

\noindent
[50] Ye. Hakobyan, V. Ter-Antonyan, "Quantum Oscillator as 1D Anyon",
quant-ph/0002069.

\noindent
[51] A. Nersessian, V. Ter-Antonyan, Phys. Atom. Nucl. {\bf 61}, 1756 (1998).

\vspace{0.3cm}

{\it Dyon--Oscillator Duality (Non-Abelian Monopole)}:

\noindent
[52] L. Mardoyan, A. Sissakian, V. Ter-Antonyan, Phys. Atom. Nucl. {\bf 61}, 1746 (1998).

\noindent
[53] L. Mardoyan, A. Sissakian, V. Ter-Antonyan, Mod. Phys. Lett. {\bf A 14}, 1303
(1999).

\end{document}